\documentclass[prd,onecolumn,nopacs,nofootinbib]{revtex4}

\usepackage{amsmath}
\usepackage{graphicx}
\usepackage{amsfonts}
\usepackage{latexsym}
\usepackage{bbold}
\usepackage{wasysym}
\usepackage{calligra}
\usepackage{float}
\usepackage{ulem}
\usepackage{inputenc}
\usepackage{xspace}
\usepackage{url}
\usepackage{epstopdf}
\usepackage{tikz}

\newtheorem{mydef}{Definition}
\newtheorem{corollary}{Corollary}

\newtheorem{theo}{Theorem}

\newcommand{\be}{\begin{equation}}
\newcommand{\ee}{\end{equation}}
\newcommand{\beq} {\begin{equation}}
\newcommand{\eeq} {\end{equation}}
\newcommand{\ba}{\begin{eqnarray}}
\newcommand{\ea}{\end{eqnarray}}
\newcommand{\D}{\Delta}
\newcommand{\A}{\alpha}

\begin{document}

	\title{Quadratic Metric-Affine Gravity: Solving for the Affine-Connection}
	
	\author{Damianos Iosifidis}
	\affiliation{Institute of Theoretical Physics, Department of Physics
		Aristotle University of Thessaloniki, 54124 Thessaloniki, Greece}
	\email{diosifid@auth.gr}
	
	\date{\today}
	\begin{abstract}
	We consider the most general 11 parameter parity even quadratic Metric-Affine Theory whose action consists of the usual Einstein-Hilbert plus the 11 quadratic terms in torsion, non-metricity as well as their mixing. By following a certain procedure and using a simple trick we are able to find the unique solution of the affine connection in terms of an arbitrary hypermomentum. Given a fairly general non-degeneracy condition our result provides the exact form of the affine connection for  all types of matter. Subsequently we compute the forms of torsion and non-metricity in terms of their sources (hypermomentum tensor) and also express the metric field equations in effectively Einstein's GR with modified source terms that depend on the hypermomentum and its derivatives. We show that in the absence of matter the Theory always reduces to GR. Finally we generalize our result and find the form of the connection for a wider class of quadratic Theories.
		
	\end{abstract}
	
	\maketitle
	
	\allowdisplaybreaks
	
	
	\tableofcontents
	
	\section{Introduction}
	\label{intro}

The geometric interpretation of gravity initiated by Einstein on the development of General Relativity (GR) is indeed an astonishing one. However, the geometric arena GR lies upon is that of Riemannian geometry \cite{eisenhart2016riemannian} which is built on the assumptions of a symmetric and metric-compatible connection. That is, a Riemannian space possesses no torsion (i.e. symmetric connection) and has vanishing non-metricity (metric-compatibility). Consequently, gravity is descried solely by the curvature and the only dynamical field is the metric tensor. Interestingly, equivalent formulations of GR can be formulated by attributing gravity to either solely torsion (teleparallel gravity \cite{aldrovandi2012teleparallel}), solely non-metricity (symmetric teleparallel \cite{nester1998symmetric,jimenez2018teleparallel}) or both (general teleparallel \cite{jimenez2019general}) by restricting the connection in a certain way. Even more intriguing is the possibility of leaving the underlying geometry free of restrictions and be in the realm of non-Riemannian geometry \cite{eisenhart2012non}. This generalized arena has all, curvature, torsion and non-metricity non-vanishing and the gravitational Theories described by non-Riemannian geometry go by the name Metric-Affine Gravity (MAG) \cite{hehl1995metric,iosifidis2019metric,gronwald1997metric}\footnote{Some recent developments in MAG and applications include \cite{iosifidis2019metric,iosifidis2019exactly,Iosifidis:2020gth,iosifidis2021perfect,percacci2020new,shimada2019metric,delhom2019ricci,iosifidis2019torsion,hohmann2020metric,Mikura:2020qhc,Jimenez:2020dpn,Ariwahjoedi:2020wmo,klemm2020einstein,vitagliano2011dynamics,iosifidis2019scale,iosifidis2021cosmology} }.

MAG is very interesting to study for many reasons. Firstly, when written in the language of differential forms it can be seen to represent a gauge Theory of gravity \cite{hehl1995metric}. Next, the modifications appearing in MAG (in comparison to GR) are due to spacetime torsion and non-metricity with both of them having a clear geometric interpretation (see for instance \cite{iosifidis2019metric}). Furthermore, an essential characteristic of MAG is the extended geometry-intrinsic structure interrelation. More precisely, in MAG one also has the notion of hypermomentum source \cite{hehl1976hypermomentum,hehl1995metric} on top of the usual energy-momentum tensor. The hypermomentum tensor is formally defined by the variation of the matter sector of the action with respect to the independent affine connection and in it it encodes the spin, dilation and shear of matter  \cite{hehl1976hypermomentum,hehl1995metric}.We see therefore that MAG represents a generalization of the well known Einstein-Cartan Theory where now apart from the spin, the dilation and shear\footnote{For an interesting possible relation among the shear current and the hadronic properties of matter see \cite{Hehl:1978cb}.} of matter (i.e. all intrinsic parts) are taken into account. 

Of course, in order to explore this Metric-Affine land the first thing we should be concerned with is the choice of the gravitational Lagrangian. Admittedly, the most obvious possibility for the latter is the generalized Einstein-Hilbert term (i.e. the full Ricci scalar) which contains non-Riemannian parts as well. However, one soon realises that this choice encounters already an issue: The invariance of the Ricci scalar under projective transformations of the connection (see for instance \cite{iosifidis2020linear}) demands an identically vanishing dilation current. As a result only matter with no dilation charge for the hypermomentum is allowed by the Theory. The most straightforward  remedy for this  problem is to supplement the Gravitational action with quadratic torsion and non-metricity invariants that would spoil the aforementioned invariance. In fact, the inclusion of these  quadratic terms is not simply a matter of convenience for breaking the projective invariance but their presence is indispensable if one wants to keep generality. Indeed, was already observed in \cite{vitagliano2011dynamics} (see also \cite{obukhov1996exact} ), from a field theoretic point of view, the dimension of the quadratic torsion and non-metricity invariants is the same with that of the Ricci scalar (being inverse length) and since there is no fundamental principle to exclude them they must appear in the Gravitational action for MAG to this order. It is known that there are $3$ pure torsion, $5$ pure non-metricity and $3$ mixed quadratic (parity even) scalars \cite{pagani2015quantum,iosifidis2019scale}. Therefore, the gravitational part of the parity even quadratic MAG action consists of the usual Einstein-Hilbert term along with the $11$ quadratic torsion and non-metricity scalars. Of course, to the aforementioned gravitational part one must also add a matter sector that will (in general) possess a non-vanishing hypermomentum\footnote{The Cosmology of the full $11$ parameter quadratic Theory in the presence of a Cosmological hyperfluid has been recently studied in \cite{iosifidis2021cosmology}.}. Then, after varying with respect to the metric and the independent affine connection the first task is to find the form of the affine connection which will subsequently reveal the expressions for torsion and non-metricity. 

Surprisingly, up to now there is no prescribed method allowing one to solve for the affine connection for the general quadratic Theory when the hypermomentum is non-vanishing. It is exactly this point we will address in this work. That is, given some general non-degeneracy conditions we will find the exact and unique solution of the connection for the full quadratic Theory. More specifically, the paper is organized as follows. Firstly, we fix notation and definitions and briefly discuss the geometric arena of MAG along with the physical content of its sources. We then consider the full $11$ parameter quadratic Lagrangian and derive the associated metric and connection field equations. Focusing on the latter and using the recent result of \cite{Iosifidis:2021ili} we find the unique and exact solution of the affine connection in terms of the hypermomentum and its contractions. Subsequently, we compute the exact forms of torsion and non-metricity for the quadratic Theory. Then, turning our attention to the metric field equations, we apply a post-Riemannian expansion and express the latter as effective Einstein's GR field equations with modified source terms due to the intrinsic properties of matter (i.e. hypermomentum). The resulting Theory serves as certain generalization of Einstein-Cartan Theory where now apart from spin, the dilation and shear of matter are taken into account, resulting in a non-vanishing non-metricity as well. Finally, generalizing the previous considerations, we consider a collection of matter  fields non-minimally coupled to the quadratic torsion and non-metricity invariants (i.e. promoting the $11$ parameters to functions) and also for this generalized action we derive the exact solution for the connection.

	\section{The setup}
We shall  consider a generalized $n-dim$  manifold endowed with a metric and an independent affine connection ($\mathcal{M}$, g, $\nabla$). We will use the definitions and notations of \cite{Iosifidis:2020gth} so we will go through the basic setup rather briefly. On this non-Riemannian manifold endowed with the metric $g_{\mu\nu}$ and an independent affine connection with components $\Gamma^{\lambda}_{\;\;\;\mu\nu}$, we define the curvature, torsion and non-metricity tensors according to
\beq
R^{\mu}_{\;\;\;\nu\alpha\beta}:= 2\partial_{[\alpha}\Gamma^{\mu}_{\;\;\;|\nu|\beta]}+2\Gamma^{\mu}_{\;\;\;\rho[\alpha}\Gamma^{\rho}_{\;\;\;|\nu|\beta]} \label{R}
\eeq
\beq
S_{\mu\nu}^{\;\;\;\lambda}:=\Gamma^{\lambda}_{\;\;\;[\mu\nu]}
\eeq
\beq
Q_{\alpha\mu\nu}:=- \nabla_{\alpha}g_{\mu\nu}
\eeq
and the deviation of the affine connection $\Gamma^{\lambda}_{\;\;\;\mu\nu}$ from the Levi-Civita one defines the so-called distortion tensor \cite{schouten1954ricci}
\begin{gather}
N^{\lambda}_{\;\;\;\;\mu\nu}:=\Gamma^{\lambda}_{\;\;\;\mu\nu}-\widetilde{\Gamma}^{\lambda}_{\;\;\;\mu\nu}=
\frac{1}{2}g^{\alpha\lambda}(Q_{\mu\nu\alpha}+Q_{\nu\alpha\mu}-Q_{\alpha\mu\nu}) -g^{\alpha\lambda}(S_{\alpha\mu\nu}+S_{\alpha\nu\mu}-S_{\mu\nu\alpha}) \label{N}
\end{gather}
where $\widetilde{\Gamma}^{\lambda}_{\;\;\;\mu\nu}$ is the usual Levi-Civita connection calculated only by the metric and its first derivatives. Once the distortion is given, torsion and non-metricity can be quite easily computed through (see for instance \cite{iosifidis2019metric})
\beq
S_{\mu\nu\alpha}=N_{\alpha[\mu\nu]}\;\;,\;\;\; Q_{\nu\alpha\mu}=2 N_{(\alpha\mu)\nu} \label{QNSN}
\eeq
Out of torsion we can construct a vector as well as a pseudo-vector. Our definitions for the torsion vector and pseudo-vector are
\beq
S_{\mu}:=S_{\mu\lambda}^{\;\;\;\;\lambda} \;\;, \;\;\;
t_{\mu}:=\epsilon_{\mu\alpha\beta\gamma}S^{\alpha\beta\gamma} 
\eeq
respectively. Note that the former is defined for any dimension while the latter only for $n=4$. Continuing with non-metricity, we define the Weyl  and the second non-metricity vector according to
\beq
Q_{\alpha}:=Q_{\alpha\mu\nu}g^{\mu\nu}\;,\;\; q_{\nu}=Q_{\alpha\mu\nu}g^{\alpha\mu}
\eeq
Finally,  from the Riemann tensor we can construct the three contractions
\beq
R_{\nu\beta}:=R^{\mu}_{\;\;\nu\mu\beta}	
\eeq
\beq
\hat{R}_{\alpha\beta}:=R^{\mu}_{\;\;\mu\alpha\beta}	
\eeq
\beq
\breve{R}^{\mu}_{\;\;\beta}:=R^{\mu}_{\;\;\nu\alpha\beta}	g^{\nu\alpha}
\eeq

As usual, the first one is the Ricci tensor (which is not symmetric in general), the second one is the homothetic curvature and the last one is the the co-Ricci tensor. Note that the first two aforementioned tensors can be formed without the use of any metric while for the latter a metric is required. As for the generalized Ricci scalar, the latter is still uniquely defined since
\beq
R:=g^{\mu\nu}R_{\mu\nu}=-g^{\mu\nu}\breve{R}_{\mu\nu} \;\; ,\;\;\; g^{\mu\nu}\hat{R}_{\mu\nu}=0
\eeq

Furthermore, let us mention that by virtue of ($\ref{N}$) each quantity can be split into its Riemannian part (i.e. computed with respect to the Levi-Civita connection) plus non-Riemannian contributions. For instance inserting the connection decomposition ($\ref{N}$) into the definition ($\ref{R}$) we obtain for the Riemann tensor \footnote{Quantities with\;  $\widetilde{}$\;  will always denote Riemannian parts  unless otherwise stated.} 
\beq
{R^\mu}_{\nu \alpha \beta} = \widetilde{R}^\mu_{\phantom{\mu} \nu \alpha \beta} + 2 \widetilde{\nabla}_{[\alpha} {N^\mu}_{|\nu|\beta]} + 2 {N^\mu}_{\lambda|\alpha} {N^\lambda}_{|\nu|\beta]} \,, \label{decomp}
\eeq
The last decomposition is very useful and we are going to be using it later on  in order to express the metric field equations of our Theory in Einstein-like form plus modified matter sources. For instance, with the use of the above the post-Riemannian expansion of the Ricci scalar reads
\begin{gather}
R=\tilde{R}+ \frac{1}{4}Q_{\alpha\mu\nu}Q^{\alpha\mu\nu}-\frac{1}{2}Q_{\alpha\mu\nu}Q^{\mu\nu\alpha}    -\frac{1}{4}Q_{\mu}Q^{\mu}+\frac{1}{2}Q_{\mu}q^{\mu}+S_{\mu\nu\alpha}S^{\mu\nu\alpha}-2S_{\mu\nu\alpha}S^{\alpha\mu\nu}-4S_{\mu}S^{\mu} \nonumber \\ +2 Q_{\alpha\mu\nu}S^{\alpha\mu\nu}+2 S_{\mu}(q^{\mu}-Q^{\mu}) +\tilde{\nabla}_{\mu}(q^{\mu}-Q^{\mu}-4S^{\mu})
\end{gather}

 We now close this section with a final definition before we turn our attention into the material content of Metric-Affine Theories.
 \begin{mydef}Consider the components $\Psi_{\alpha\mu\nu}$ of a rank-3 tensor field. We define the $1^{st}$, $2^{nd}$ and $3^{rd}$ contractions of $\Psi_{\alpha\mu\nu}$ according to
	\beq
	\Psi^{(1)}_{\mu}:=\Psi_{\alpha\beta\mu}g^{\alpha\beta}\;\;, \;\; \Psi^{(2)}_{\mu}:=\Psi_{\alpha\mu\beta}g^{\alpha\beta}\;\;, \;\; \Psi^{(3)}_{\mu}:=\Psi_{\mu\alpha\beta}g^{\alpha\beta}
	\eeq
	respectively. 
\end{mydef}
Let us now move on and discuss some basic physical aspects of MAG.

	\section{The sources of 
		MAG}
\subsection{Canonical and Metrical Energy Momentum and Hypermomentum Tensors}
As we have already pointed out, in the holonomic\footnote{In an anholonomic description we have three independent fields $g_{ab}, \vartheta^{b}$ and $\Gamma^{a}_{\;\; b}$, that is the tangent space metric, the co-frame and the spin connection. However, the metric field equations are redundant in this case and therefore we have only two independent fields in this case as well (see \cite{hehl1995metric}) } description of MAG  one starts  with an independent fields affine connection $\Gamma^{\lambda}_{\;\;\;\mu\nu}$ along with the metric $g_{\mu\nu}$. The field equations of a given MAG Theory are obtained by varying the total action independently with respect to those fields. Then, the variations of the matter part of the actions would be the sources of Gravity. Let $\mathcal{L}_{m}$ be the matter Lagrangian of the Theory. As usual we define the energy-momentum tensor of matter by the metric variation of the matter sector, viz. 
\beq
T^{\alpha\beta}:=+\frac{2}{\sqrt{-g}}\frac{\delta(\sqrt{-g} \mathcal{L}_{M})}{\delta g_{\alpha\beta}}
\eeq
On the other hand, since we now have an independent affine connection on which the matter part can just as well depend upon, we define the variation of the latter with respect to the affine connection
\beq
\Delta_{\lambda}^{\;\;\;\mu\nu}:= -\frac{2}{\sqrt{-g}}\frac{\delta ( \sqrt{-g} \mathcal{L}_{M})}{\delta \Gamma^{\lambda}_{\;\;\;\mu\nu}}
\eeq
 Note that the hypermomentum can be split into its three  irreducible pieces of spin, dilation and shear according to\footnote{Each of these irreducible pieces forms an invariant subgroup under rotations.}
\beq
\Delta_{\mu\nu\alpha}=\tau_{\mu\nu\alpha}+\frac{1}{n}\Delta_{\alpha} g_{\mu\nu}+\hat{\Delta}_{\mu\nu\alpha}
\eeq
where $\tau_{\mu\nu\alpha}:=\Delta_{[\mu\nu]\alpha}$ is the spin part, $\Delta_{\alpha}:=\Delta_{\mu\nu\alpha}g^{\mu\nu}$ the dilation (trace) and $\hat{\Delta}_{\mu\nu\alpha}=\Delta_{(\mu\nu)\alpha}-\frac{1}{n}\Delta_{\alpha} g_{\mu\nu}$ the shear (symmetric traceless part).
The aforementioned sources are subject to the conservation law
\beq
\sqrt{-g}(2 \tilde{\nabla}_{\mu}T^{\mu}_{\;\;\alpha}-\Delta^{\lambda\mu\nu}R_{\lambda\mu\nu\alpha})+\hat{\nabla}_{\mu}\hat{\nabla}_{\nu}(\sqrt{-g}\Delta_{\alpha}^{\;\;\mu\nu})+2S_{\mu\alpha}^{\;\;\;\;\lambda}\hat{\nabla}_{\nu}(\sqrt{-g}\Delta_{\lambda}^{\;\;\;\mu\nu})=0 \label{ccc}
\eeq
which serves as the generalization of the conservation law of energy-momentum for matter with microstructure (see)
Now we have all the ingredients needed to study model building and more specifically quadratic MAG Theories.

\section{Full Quadratic Theory}

In order to set up a Metric-Affine Theory, the first thing  one has to specify is the underlying gravitational action that is going to be used in constructing the Theory. The most obvious possibility is to replace the usual (Riemannian) Einstein-Hilbert term with the full Ricci scalar, namely $\tilde{R}\rightarrow R$ and consider the gravitational action
\beq
S_{EH}[g, \Gamma]=\frac{1}{2 \kappa}\int d^{n}x \sqrt{-g}R \label{EH}
\eeq
as the representative one for MAG. Even though this is indeed a good choice once the matter part of the action does not depend on the connection (i.e. for matter with $\Delta_{\alpha\mu\nu}=0$), in the sense that it gives Einstein's field equations, problems set in when matter couples to the connection. It is known that the problem lies on the projective invariance of the Ricci scalar (see for instance) which then implies a vanishing dilation current i.e. $\Delta^{\mu}=0$ once the connection field equations are imposed. It is therefore necessary, from a physical point of view, to consider actions that  go beyond the mere generalization of the non-Riemannian Einstein-Hilbert term and break the projective invariance. Interestingly enough, the above reason is not the only one that demands a modification to the above action. There is yet another more fundamental reason enforcing us to extend $\ref{EH}$  that has to do with dimensional analysis. Indeed, as it was already noted in  \cite{obukhov1996exact} (see also \cite{vitagliano2011dynamics}) the dimension of the Ricci scalar is inverse squared length [$L^{-2}$]. However, since the connection has dimensions of inverse length, quadratic torsion and non-metricity invariants have the same dimensions [$L^{-2}$] with the Ricci scalar. Therefore, from an effective field Theory perspective these terms may as well supplement the Ricci scalar, being of the same dimension. It turns out that to quadratic order there are $3$ pure torsion, $5$ pure non-metricity and $3$ mixed scalars (see for instance \cite{pagani2015quantum,iosifidis2019scale}). Therefore, the most general parity even quadratic MAG action in arbitrary dimension $n$ reads 
\begin{gather}
S_{quad}
=\frac{1}{2 \kappa}\int d^{n}x \sqrt{-g} \Big[   
b_{1}S_{\alpha\mu\nu}S^{\alpha\mu\nu} +
b_{2}S_{\alpha\mu\nu}S^{\mu\nu\alpha} +
b_{3}S_{\mu}S^{\mu} \nonumber \\
a_{1}Q_{\alpha\mu\nu}Q^{\alpha\mu\nu} +
a_{2}Q_{\alpha\mu\nu}Q^{\mu\nu\alpha} +
a_{3}Q_{\mu}Q^{\mu}+
a_{4}q_{\mu}q^{\mu}+
a_{5}Q_{\mu}q^{\mu} \nonumber \\
+c_{1}Q_{\alpha\mu\nu}S^{\alpha\mu\nu}+
c_{2}Q_{\mu}S^{\mu} +
c_{3}q_{\mu}S^{\mu} \label{quad}
\Big] 
\end{gather}
where the $a_{i}'s$, $b_{j}'s$ and $c_{k}'s$ are dimensionless coupling constants. Of course to the above one could also add scalars constructed by the covariant derivatives of torsion and non-metricity, like for instance $g^{\mu\nu}\nabla_{\mu}Q_{\nu}$. However, one can always apply a post-Riemannian expansion to such terms ending up with total derivative terms plus some combinations of the above-included quadratic torsion and non-metricity invariance\footnote{For a detailed example see  \cite{vitagliano2011dynamics}.}. As a result the above $11$ parameter action is indeed the most general  quadratic one (and parity even). Note that in \cite{vitagliano2011dynamics} the connection solution of the action was found but only when the $3$ quadratic torsion terms where included. Our intention here is to generalize these results and find the connection solution for the full quadratic action $(\ref{quad})$.

\section{The  Quadratic Parity Even MAG Theory}
As we already mentioned the main intention of this work is to analyze the full $11$-parameter quadratic MAG Theory including all possible, parity even  terms in torsion and non-metricity in addition to the Einstein-Hilbert term. Adding a matter part as well, our full action reads
\begin{gather}
S[g, \Gamma, \Phi]
=\frac{1}{2 \kappa}\int d^{n}x \sqrt{-g} \Big[  R+ 
b_{1}S_{\alpha\mu\nu}S^{\alpha\mu\nu} +
b_{2}S_{\alpha\mu\nu}S^{\mu\nu\alpha} +
b_{3}S_{\mu}S^{\mu} \nonumber \\
a_{1}Q_{\alpha\mu\nu}Q^{\alpha\mu\nu} +
a_{2}Q_{\alpha\mu\nu}Q^{\mu\nu\alpha} +
a_{3}Q_{\mu}Q^{\mu}+
a_{4}q_{\mu}q^{\mu}+
a_{5}Q_{\mu}q^{\mu} \nonumber \\
+c_{1}Q_{\alpha\mu\nu}S^{\alpha\mu\nu}+
c_{2}Q_{\mu}S^{\mu} +
c_{3}q_{\mu}S^{\mu} 
\Big] +S_{M}[g, \Gamma, \Phi]
\end{gather}
In the above, the first, second and third lines contain (apart from the usual Einstein-Hilbert term) the pure torsion, pure non-metricity and mixed quadratic parity even parts\footnote{In \cite{obukhov1997irreducible} to this quadratic Lagrangian, a homothetic curvature squared term was added $\hat{R}_{\mu\nu}\hat{R}^{\mu\nu}$. This term however is of dimension [$L^{-4}$] and its inclusion is not along the lines of effective field Theory approach. We shall therefore consider no such terms in our study here.}. Let us highlight that the matter action we are considering here can (and will) in general depend on the connection which will result in a non-vanishing hypermomentum tensor. Varying the above action independently with respect to the metric and the affine connection, after some long calculations we obtain the field equations

\textbf{g-Variation}
	\beq
R_{(\mu\nu)}-\frac{R}{2}g_{\mu\nu}-\frac{\mathcal{L}_{2}}{2}g_{\mu\nu}+\frac{1}{\sqrt{-g}}\hat{\nabla}_{\alpha}\Big( \sqrt{-g}(W^{\alpha}_{\;\;\;(\mu\nu)}+\Pi^{\alpha}_{\;\;\;(\mu\nu)})\Big) +A_{(\mu\nu)}+B_{(\mu\nu)}+C_{(\mu\nu)}=\kappa T_{\mu\nu} \label{metricf}
\eeq
where we have abbreviated
\beq
\hat{\nabla}_{\mu}:=2S_{\mu}-\nabla_{\mu}	
\eeq
\beq
W^{\alpha}_{\;\;(\mu\nu)}=2 a_{1}Q^{\alpha}_{\;\;\mu\nu}+2 a_{2}Q_{(\mu\nu)}^{\;\;\;\;\alpha}+(2 a_{3}Q^{\alpha}+a_{5}q^{\alpha})g_{\mu\nu}+(2 a_{4}q_{(\mu} + a_{5}Q_{(\mu})\delta^{\alpha}_{\nu)}
\eeq
\beq
\Pi^{\alpha\mu\nu} = c_{1}S^{\alpha\mu\nu}+c_{2}g^{\mu\nu}S^{\alpha}+c_{3}g^{\alpha\mu}S^{\nu}
\eeq

\beq
A_{\mu\nu}=a_{1}(Q_{\mu\alpha\beta}Q_{\nu}^{\;\;\alpha\beta}-2 Q_{\alpha\beta\mu}Q^{\alpha\beta}_{\;\;\;\;\nu})-a_{2}Q_{\alpha\beta(\mu}Q^{\beta\alpha}_{\;\;\;\;\nu)}
+a_{3}(Q_{\mu}Q_{\nu}-2 Q^{\alpha}Q_{\alpha\mu\nu})-a_{4}q_{\mu}q_{\nu}-a_{5}q^{\alpha}Q_{\alpha\mu\nu}
\eeq	
\beq
B_{\mu\nu}=b_{1}(2S_{\nu\alpha\beta}S_{\mu}^{\;\;\;\alpha\beta}-S_{\alpha\beta\mu}S^{\alpha\beta}_{\;\;\;\;\nu})-b_{2}S_{\nu\alpha\beta}S_{\mu}^{\;\;\;\beta\alpha}+b_{3}S_{\mu}S_{\nu} 
\eeq
\beq
C_{\mu\nu}=\Pi_{\mu\alpha\beta}Q_{\nu}^{\;\;\;\alpha\beta}	-( c_{1}S_{\alpha\beta\nu}Q^{\alpha\beta}_{\;\;\;\;\mu}+c_{2}S^{\alpha}Q_{\alpha\mu\nu}+c_{3}S^{\alpha}Q_{\mu\nu\alpha})=c_{1}(Q_{\mu}^{\;\;\;\alpha\beta}S_{\nu\alpha\beta}-S_{\alpha\beta\mu}Q^{\alpha\beta}_{\;\;\;\;\nu})+c_{2}(S_{\mu}Q_{\nu}-S^{\alpha}Q_{\alpha\mu\nu})
\eeq

\textbf{$\Gamma$-Variation}

\begin{gather}
\left( \frac{Q_{\lambda}}{2}+2 S_{\lambda}\right)g^{\mu\nu}-Q_{\lambda}^{\;\;\mu\nu}-2 S_{\lambda}^{\;\;\mu\nu}+\left( q^{\mu} -\frac{Q^{\mu}}{2}-2 S^{\mu}\right)\delta_{\lambda}^{\nu}+4 a_{1}Q^{\nu\mu}_{\;\;\;\;\lambda}+2 a_{2}(Q^{\mu\nu}_{\;\;\;\;\lambda}+Q_{\lambda}^{\;\;\;\mu\nu})+2 b_{1}S^{\mu\nu}_{\;\;\;\;\lambda} \nonumber \\
+2 b_{2}S_{\lambda}^{\;\;\;[\mu\nu]}+c_{1}\Big( S^{\nu\mu}_{\;\;\;\;\lambda}-S_{\lambda}^{\;\;\;\nu\mu}+Q^{[\mu\nu]}_{\;\;\;\;\;\lambda}\Big)+\delta_{\lambda}^{\mu}\Big( 4 a_{3}Q^{\nu}+2 a_{5}q^{\nu}+2 c_{2}S^{\nu}\Big)+\delta_{\lambda}^{\nu}\Big(  a_{5}Q^{\mu}+2 a_{4}q^{\mu}+ c_{3}S^{\mu}\Big) \nonumber \\
+g^{\mu\nu}\Big(a_{5} Q_{\lambda}+2 a_{4}q_{\lambda}+c_{3}S_{\lambda} \Big)+\Big( c_{2} Q^{[\mu}+ c_{3}q^{[\mu}+2 b_{3}S^{[\mu}\Big) \delta^{\nu]}_{\lambda}  =\kappa \Delta_{\lambda}^{\;\;\;\mu\nu} \label{G}
\end{gather}
The metric field equations ($\ref{metricf}$) are the analogue of the Einstein field equations where now apart from the explicit additional terms of torsion and non-metricity, both the Ricci tensor and scalar contain post-Riemannian pieces as well. The exact deviation from the usual field equations of GR will be clear after we solve for the affine connection and substitute all post-Riemannian expansions back in ($\ref{metricf}$). This will be clearly illustrated in what follows. Before we move on to solve for the connection let us elaborate a little bit in the metric field equations. Taking the trace of the latter we obtain
	\beq
\Big( 1-\frac{n}{2} \Big) R+\Big( 1-\frac{n}{2} \Big) \mathcal{L}_{2}-\frac{1}{\sqrt{-g}}\tilde{\nabla}_{\alpha}\Big(\sqrt{-g}(\Pi^{\alpha}+W^{\alpha})\Big)=\kappa T \label{trace}
\eeq
where
\beq
\Pi^{\alpha}:=\Pi^{\alpha}_{\;\;\mu\nu}g^{\mu\nu}=(c_{1}+n c_{2}+c_{3})S^{\alpha}
\eeq
\beq
W^{\alpha}:=W^{\alpha}_{\;\;\mu\nu}g^{\mu\nu}=(2 a_{1}+2 n a_{3}+a_{5})Q^{\alpha}+(2 a_{2} +2 a_{4}+n a_{5})q^{\alpha}
\eeq
Expanding out the Ricci scalar into its Riemannian part plus non-Riemannian contributions, equation ($\ref{trace}$) becomes
\begin{gather}
\tilde{R}
+	\Big( a_{1}+\frac{1}{4}\Big)  Q_{\alpha\mu\nu}Q^{\alpha\mu\nu} +
\Big( a_{2}-\frac{1}{2}\Big)Q_{\alpha\mu\nu}Q^{\mu\nu\alpha} +
\Big( a_{3}-\frac{1}{4}\Big)Q_{\mu}Q^{\mu}+
a_{4}q_{\mu}q^{\mu}+
\Big( a_{5}+\frac{1}{2}\Big)Q_{\mu}q^{\mu} \nonumber \\
+(b_{1}+1)S_{\alpha\mu\nu}S^{\alpha\mu\nu} +
(b_{2}-2)S_{\alpha\mu\nu}S^{\mu\nu\alpha} +
(b_{3}-4)S_{\mu}S^{\mu}
+	(c_{1}+2)Q_{\alpha\mu\nu}S^{\alpha\mu\nu}+
(c_{2}-2)Q_{\mu}S^{\mu} +
(c_{3}+2)q_{\mu}S^{\mu} \nonumber \\
+\frac{1}{\sqrt{-g}}\widetilde{\nabla}_{\mu}\Big[ \sqrt{-g}\Big( (2a_{2}+ 2 a_{4}+n a_{5}+1)q^{\mu}+(2 a_{1}+2 n a_{3}+a_{5}-1)Q^{\mu}+(c_{1}+n c_{2}+c_{3}-4)S^{\mu} \Big) \Big] =\kappa T \label{effEin}
\end{gather}
Note that for the parameter choice  $a_{1}=-a_{3}=-1/4$, $a_{2}=-a_{5}=1/2$, $a_{4}=0$,  $b_{1}=-1$, $b_{2}=1$, $b_{3}=4$ , $c_{1}=-c_{2}=c_{3}=-2$ all non-Riemannian pieces of torsion and non-metricity disappear and $(\ref{effEin})$ is the same with the trace of Einstein's field equations of GR. The same holds true for the Ricci tensor and essentially the Theory is GR for this parameter choice. Of course this is no surprise, since choosing the parameters this way all the additional quadratic torsion and  non-metricity terms exactly cancel the corresponding ones coming from the post Riemannian expansion of the Ricci scalar yielding just $\tilde{R}$ in the end. However, this is extreme fine tuning and in order to keep our discussion as general as possible we shall not impose any \textit{a priori} relation among the couplings of the Theory.

\section{Solving for the affine connection}
We shall now pursue to solve for the affine connection and derive its exact form for the quadratic Theory of the previous section. Even though the connection field equations $(\ref{G})$ are fairly complicated as we show below, using a certain result and also performing a simple trick the general solution can be obtained quite easily. We are going to use the following Theorem which we state here without proof. The detailed proof can be found in \cite{Iosifidis:2021ili}.

\begin{theo}
	Consider the $15$ parameter linear tensor equation 
		\beq
	\alpha_{1}N_{\alpha\mu\nu}+\alpha_{2}N_{\nu\alpha\mu}+\alpha_{3}N_{\mu\nu\alpha}+\alpha_{4}N_{\alpha\nu\mu}+\alpha_{5}N_{\nu\mu\alpha}+\alpha_{6}N_{\mu\alpha\nu}+\sum_{i=1}^{3}\Big( \alpha_{7i}N^{(i)}_{\mu}g_{\alpha\nu}+\alpha_{8i}N^{(i)}_{\nu}g_{\alpha\mu}+\alpha_{9i}N^{(i)}_{\alpha}g_{\mu\nu} \Big)=B_{\alpha\mu\nu} \label{eq4}
	\eeq
	 where $N_{\alpha\mu\nu}$ are the components of the unknown tensor field, $B_{\alpha\mu\nu}$ the components of a given (known) tensor field and $\alpha_{i}$, $i=1,2,...,6$ are scalars. Define the matrices\footnote{The elements $\gamma_{ij}$ are linear combinations of the parameters $a_{i}$ and their exact relations are given in the appendix.}
	\begin{equation}
	\Gamma := 
	\begin{pmatrix}
	\gamma_{11} & \gamma_{12} & \gamma_{13}  \\
	\gamma_{21} & \gamma_{22} &  \gamma_{23} \\
	\gamma_{31} & \gamma_{32} &  \gamma_{33}  \\
	\end{pmatrix}
	\end{equation}
	and
	\begin{equation}
A := 
\begin{pmatrix}
\A_{1} & \A_{2} & \A_{3} & \A_{4} & \A_{5} & \A_{6} \\
\A_{3} & \A_{1} &  \A_{2} & \A_{5} & \A_{6} & \A_{4} \\
\A_{2} & \A_{3} &  \A_{1} & \A_{6} & \A_{4} & \A_{5} \\
\A_{4} & \A_{6} &  \A_{5} & \A_{1} & \A_{3} & \A_{2} \\
\A_{5} & \A_{4} &  \A_{6} & \A_{2} & \A_{1} & \A_{3} \\
\A_{6} & \A_{5} &  \A_{4} & \A_{3} & \A_{2} & \A_{1} \\
\end{pmatrix}
\end{equation}
	Then, given that both of the above matrices are non-singular, the unique  solution to $(\ref{eq4})$
	reads
	\beq
	N_{\alpha\mu\nu}=\tilde{a}_{11}	B_{\alpha\mu\nu}+\tilde{a}_{12}\hat{B}_{\nu\alpha\mu}+\tilde{a}_{13}\hat{B}_{\mu\nu\alpha}+\tilde{a}_{14}\hat{B}_{\alpha\nu\mu}
	+\tilde{a}_{15} \hat{B}_{\nu\mu\alpha}+\tilde{a}_{16} \hat{B}_{\mu\alpha\nu} \label{co2}
	\eeq
	where $\tilde{a}_{1j}$ $j=1,2,...,6$ are the first row elements of the inverse matrix $A^{-1}$ and
	\beq
	\hat{B}_{\alpha\mu\nu}={B}_{\alpha\mu\nu}-\sum_{i=1}^{3}\sum_{j=1}^{3}\Big( \alpha_{7i}\tilde{\gamma}_{ij}B^{(j)}_{\mu}g_{\alpha\nu}+\alpha_{8i}\tilde{\gamma}_{ij}B^{(j)}_{\nu}g_{\alpha\mu}+\alpha_{9i}\tilde{\gamma}_{ij}B^{(j)}_{\alpha}g_{\mu\nu}\Big) \label{Bhat}
	\eeq
		
\end{theo}
An immediate consequence of the above result is the following. 

\begin{corollary}
If $B_{\alpha\mu\nu}=0$ and both matrices $A$ and $\Gamma$  are not singular then $N_{\alpha\mu\nu}=0$ is the unique solution of $(\ref{eq4})$.
\end{corollary}

Let us now go back into the field equations of the quadratic Theory and in particular consider the connection field equations ($\ref{G}$).

A clever trick now will allow us to solve for the connection. The crucial step  is to use equations ($\ref{QNSN}$) and also their  contractions (see appendix) in order eliminate torsion and non-metricity and  express everything in terms of the distortion tensor and its associated vectors. After some lengthy but rather straightforward algebra we finally arrive at	
	\begin{gather}
( 4a_{1}+b_{1}-c_{1})N_{\alpha\mu\nu}+\Big( -1+2 a_{2}+\frac{c_{1}+b_{2}}{2}\Big)N_{\nu\alpha\mu}+\Big( -1+2 a_{2}+\frac{c_{1}+b_{2}}{2}\Big)N_{\mu\nu\alpha}\nonumber \\
+(2 a_{2}-b_{1}+c_{1})N_{\alpha\nu\mu}+\Big( 2 a_{2}-\frac{b_{2}}{2}\Big)
N_{\nu\mu\alpha}+\Big( 4 a_{1}-\frac{b_{2}}{2}-c_{1}\Big) N_{\mu\alpha\nu} \nonumber \\
+\left(2 a_{5}+c_{2}-\frac{b_{3}+c_{3}}{2}\right)g_{\nu\alpha} N^{(1)}_{\mu}+\left(2 a_{4}+\frac{b_{3}}{2}+c_{3}\right)g_{\nu\alpha} N^{(2)}_{\mu}+\left( 1+ 2a_{4}+\frac{c_{3}}{2}\right) g_{\nu\alpha}  N^{(3)}_{\mu} \nonumber \\
+\left( 8 a_{3}- 2 c_{2}+\frac{b_{3}}{2}\right) g_{\mu\alpha}N^{(1)}_{\nu}+\left( 2 a_{5}+c_{2}-\frac{c_{3}+b_{3}}{2}\right) g_{\mu\alpha}N^{(2)}_{\mu}+\left( 2 a_{5}-\frac{c_{3}}{2}\right) g_{\mu\alpha}N^{(3)}_{\nu} \nonumber \\
+\left( 2 a_{5}-\frac{c_{3}}{2}\right)g_{\mu\nu} N^{(1)}_{\alpha}+\left( 2 a_{4}+\frac{c_{3}}{2}+1\right) g_{\mu\nu}N^{(2)}_{\alpha}+2 a_{4}g_{\mu\nu}N^{(3)}_{\alpha}
=\kappa \Delta_{\alpha\mu\nu} \label{Neq}
\end{gather}
Now we see that the latter equation falls in the case of $Theorem-1$ with the obvious identifications among the parameters $\alpha_{i}$ and $a_{i}, b_{j}, c_{k}$ etc. Therefore, in applying $Theorem-1$ to the latter and setting $B_{\alpha\mu\nu}=\kappa \Delta_{\alpha\mu\nu}$ we solve for the distortion
	\beq
N_{\alpha\mu\nu}=\tilde{a}_{11}	B_{\alpha\mu\nu}+\tilde{a}_{12}\hat{B}_{\nu\alpha\mu}+\tilde{a}_{13}\hat{B}_{\mu\nu\alpha}+\tilde{a}_{14}\hat{B}_{\alpha\nu\mu}
+\tilde{a}_{15} \hat{B}_{\nu\mu\alpha}+\tilde{a}_{16} \hat{B}_{\mu\alpha\nu} \label{co2}
\eeq
where 
\beq
\hat{B}_{\alpha\mu\nu}={B}_{\alpha\mu\nu}-\sum_{i=1}^{3}\sum_{j=1}^{3}\Big( \alpha_{7i}\tilde{\gamma}_{ij}B^{(j)}_{\mu}g_{\alpha\nu}+\alpha_{8i}\tilde{\gamma}_{ij}B^{(j)}_{\nu}g_{\alpha\mu}+\alpha_{9i}\tilde{\gamma}_{ij}B^{(j)}_{\alpha}g_{\mu\nu}\Big) \label{Bhat}
\eeq	
where we have assumed that the parameters of the Quadratic Theory are totally independent\footnote{In fact, if one wants to keep fair generality this is the way to go. If there were certain relations among the parameters these would signal some invariance under connection transformations which, in turn, would then impose certain constraints on the matter fields (see \cite{iosifidis2020linear}). It is therefore crucial, also from a physical point of view, not to impose any relations among the parameters of the Theory. } and as a result $det(A)\neq 0$ and $\det(\Gamma)\neq 0$ which imply that indeed ($\ref{co2}$) is the unique solution of ($\ref{Neq}$).
As a result, in the general non-degenerate case when there are no relations among the parameters, the general solution of the latter equation reads
\beq
N_{\alpha\mu\nu}=\kappa\Big( \tilde{a}_{11}	\hat{\Delta}_{\alpha\mu\nu}+\tilde{a}_{12}\hat{\Delta}_{\nu\alpha\mu}+\tilde{a}_{13}\hat{\Delta}_{\mu\nu\alpha}+\tilde{a}_{14}\hat{\Delta}_{\alpha\nu\mu}
+\tilde{a}_{15} \hat{\Delta}_{\nu\mu\alpha}+\tilde{a}_{16} \hat{\Delta}_{\mu\alpha\nu}\Big) \label{co2}
\eeq
where 
\beq
\hat{\Delta}_{\alpha\mu\nu}=\Delta_{\alpha\mu\nu}-\sum_{i=1}^{3}\sum_{j=1}^{3}\Big( \alpha_{7i}\tilde{\gamma}_{ij}\Delta^{(j)}_{\mu}g_{\alpha\nu}+\alpha_{8i}\tilde{\gamma}_{ij}\Delta^{(j)}_{\nu}g_{\alpha\mu}+\alpha_{9i}\tilde{\gamma}_{ij}\Delta^{(j)}_{\alpha}g_{\mu\nu}\Big) \label{Dhat}
\eeq
and $\tilde{\gamma}_{ij}$ are the first row elements of the inverse matrix $\Gamma^{-1}$. The relations of the $\gamma_{ij}'s$ with the parameters of the quadratic Theory are given in the appendix.
 Having solved for the distortion we then substitute this form back in the connection decomposition and find the exact expression for the affine connection in Quadratic MAG Theories, which reads 
\beq
\Gamma^{\lambda}_{\;\;\;\mu\nu}=\tilde{\Gamma}^{\lambda}_{\;\;\;\mu\nu}+\kappa\Big( \tilde{a}_{11}	\hat{\Delta}_{\alpha\mu\nu}+\tilde{a}_{12}\hat{\Delta}_{\nu\alpha\mu}+\tilde{a}_{13}\hat{\Delta}_{\mu\nu\alpha}+\tilde{a}_{14}\hat{\Delta}_{\alpha\nu\mu}
+\tilde{a}_{15} \hat{\Delta}_{\nu\mu\alpha}+\tilde{a}_{16} \hat{\Delta}_{\mu\alpha\nu}\Big) 
\eeq
with this we can then easily find the forms of torsion and non-metricity, sourced by arbitrary hypermomentum, as
\beq
S_{\mu\nu\alpha}=\kappa\Big[ (\tilde{a}_{11}-\tilde{a}_{14})\hat{\D}_{\alpha[\mu\nu]}+(\tilde{a}_{13}-\tilde{a}_{15})\hat{\D}_{[\mu\nu]\alpha}+(\tilde{a}_{16}-\tilde{a}_{12})\hat{\D}_{[\mu|\alpha|\nu]}\Big]
\eeq
and
\beq
Q_{\alpha\mu\nu}=2 \kappa \Big[ (\tilde{a}_{11}+\tilde{a}_{16})\hat{\D}_{(\mu\nu)\alpha}+(\tilde{a}_{12}+\tilde{a}_{15})\hat{\Delta}_{\alpha(\mu\nu)}+(\tilde{a}_{13}+\tilde{a}_{14})\hat{\Delta}_{(\mu|\alpha|\nu)}\Big]
\eeq
respectively. These last two expressions show how the sources (hypermomentum) excite the non-Riemannian degrees of freedom. It is interesting to note that when $\tilde{a}_{11}=\tilde{a}_{14}$, $\tilde{a}_{13}=\tilde{a}_{15}$ and $\tilde{a}_{16}=\tilde{a}_{12}$, torsion vanishes for arbitrary hypermomentum, while for $\tilde{a}_{11}=-\tilde{a}_{16}$, $\tilde{a}_{12}=-\tilde{a}_{15}$ and $\tilde{a}_{13}=-\tilde{a}_{14}$ spacetime non-metricity identically vanishes, again without any assumption about the sources. Note also that there are no parameter values for which both $S_{\mu\nu\alpha}=0$ and $Q_{\alpha\mu\nu}=0$ at the same time.\footnote{This is tightly connected with the fact that there is no transformation of the affine connection that can simultaneously transform to zero both torsion and non-metricity. For a discussion of this aspect see \cite{iosifidis2020linear}.} It also immediately follows from $Corollary$-$1$ that when the hypermomentum sources are switched-off, the Theory reduces to GR. Therefore, we have:
\begin{corollary}
 In vacuum (or more generally when the matter decouples from the connection) the quadratic Theory is indistinguishable from General Relativity as in Einstein-Cartan case. 
\end{corollary}

 Let us now examine two characteristic cases for the hypermomentum tensor.
\subsection{Totally Antisymmetric Hypermomentum (Purely Fermionic Matter)}
For purely fermionic matter sectors\footnote{Of course one can always include additional matter fields that have no dependence on the connection whatsoever. These, having a vanishing hypermomentum, do not alter the forms of torsion and non-metricity in any way.}  the hypermomentum is totally antisymmetric and it is given by the spin tensor, according to
\beq
\Delta_{\alpha\mu\nu}=\Delta_{[\alpha\mu\nu]}
\eeq
In this case the torsion tensor becomes
\beq
S_{\alpha\mu\nu}=\kappa\Big(\tilde{a}_{11}+\tilde{a}_{12}+\tilde{a}_{13}-\tilde{a}_{14}-\tilde{a}_{15}-\tilde{a}_{16}\Big)\Delta_{\alpha\mu\nu}
\eeq
while non-metricity identically vanishes
\beq
Q_{\alpha\mu\nu}\equiv 0
\eeq
Therefore, as expected we confirm that fermions cannot excite non-metricity and only induce spacetime torsion that is totally antisymmetric (i.e. only the $t^{\mu}$ part of the full tensor is excited).

\subsection{Totally Symmetric Hypermomentum}
In the case of totally symmetric hypermomentum, that is $\Delta_{\alpha\mu\nu}=\Delta_{(\alpha\mu\nu)}$, not surprisingly the roles of torsion and non-metricity are reversed. In particular, for such a matter configuration, torsion vanishes identically
\beq
S_{\alpha\mu\nu} \equiv 0
\eeq
while non-metricity contains only the totally symmetric part
\beq
Q_{\alpha\mu\nu}=\kappa\Big(\tilde{a}_{11}+\tilde{a}_{12}+\tilde{a}_{13}+\tilde{a}_{14}+\tilde{a}_{15}+\tilde{a}_{16}\Big)\Delta_{\alpha\mu\nu}
\eeq
We should mention however that even though a totally antisymmetric hypermomentum tensor possesses some  physical substance (i.e fermionic matter), there are not (as yet) known examples of fundamental particles carrying a totally symmetric hypermomentum.

\section{Effective GR with hypermomentum contributions}
Having successfully solved for the affine connection in terms of the hypermomentum sources we can then consider the full post-Riemannian expansion of the metric field equations ($\ref{metricf}$). After some lengthy calculations (see appendix for details) we finally find
\begin{gather}
	\Big( \tilde{R}_{\mu\nu}-\frac{\tilde{R}}{2}g_{\mu\nu}\Big)=\kappa T_{\mu\nu} -\tilde{\nabla}_{\alpha}N^{\alpha}_{\;\;\;(\mu\nu)}+\tilde{\nabla}_{(\nu}N^{(2)}_{\mu)}-N^{(2)}_{\lambda}N^{\lambda}_{\;\;\;(\mu\nu)}+N^{\alpha}_{\;\;\;\lambda(\nu}N^{\lambda}_{\;\;\;\mu)\alpha}\nonumber \\
+\frac{1}{2}g_{\mu\nu}\Big[ \tilde{\nabla}_{\alpha}( N^{(3)\alpha}-N^{(2)\alpha})+N^{(3)}_{\alpha}N^{(2)\alpha}-N_{\alpha\beta\gamma}N^{\beta\gamma\alpha} \Big] \nonumber \\
+\frac{g_{\mu\nu}}{2} \Big[ \Big( \frac{b_{1}}{2}+ 2 a_{1}-c_{1}\Big) N_{\alpha\beta\gamma}N^{\alpha\beta\gamma}+	\Big( -\frac{b_{1}}{2}+  a_{2}+c_{1}\Big)N_{\alpha\beta\gamma}N^{\alpha\gamma\beta} \nonumber \\
+	\Big( \frac{b_{2}}{2}+ 2 a_{2}+c_{1}\Big)N_{\alpha\beta\gamma}N^{\beta\gamma\alpha}+	\Big( -\frac{b_{2}}{2}+ 2 a_{1}+a_{2}-c_{1}\Big)N_{\mu\nu\alpha}N^{\nu\mu\alpha} \nonumber \\
+\Big( \frac{b_{3}}{4}+4 a_{3}-c_{2} \Big)N^{(1)}_{\alpha}N^{(1)}_{\beta}g^{\alpha\beta}+\Big( \frac{b_{3}}{4}+a_{4}+\frac{c_{3}}{4} \Big)N^{(2)}_{\alpha}N^{(2)}_{\beta}g^{\alpha\beta}+a _{4}N^{(3)}_{\mu}N^{(3)}_{\nu}g^{\mu\nu} \nonumber \\
+\Big( 2 a_{5}-\frac{c_{3}}{2}+c_{2}-\frac{b_{3}}{2}\Big) N^{(1)}_{\alpha}N^{(2)}_{\beta}g^{\alpha\beta}	+\Big( 2 a_{4}+\frac{c_{3}}{2}\Big)N^{(2)}_{\alpha}N^{(3)}_{\beta}g^{\alpha\beta}+\Big( 2 a_{5}-\frac{c_{3}}{2}\Big) N^{(1)}_{\alpha}N^{(3)}_{\beta}g^{\alpha\beta} \Big] \nonumber \\
- \Big( 4 a_{1}-\frac{c_{1}}{2}\Big)\frac{1}{\sqrt{-g}}\hat{\nabla}_{\beta}\Big(\sqrt{-g}g^{\alpha\beta} N_{(\mu\nu)\alpha}\Big)-\Big( 2 a_{2}+\frac{c_{1}}{2}\Big)\frac{1}{\sqrt{-g}}\hat{\nabla}_{\beta}\Big(\sqrt{-g}g^{\alpha\beta} N_{(\mu|\alpha|\nu)}\Big)-2 a_{2}\frac{1}{\sqrt{-g}}\hat{\nabla}_{\beta}\Big(\sqrt{-g}g^{\alpha\beta}N_{\alpha(\mu\nu)}\Big) \nonumber \\
-\frac{1}{\sqrt{-g}}\hat{\nabla}_{\alpha}\left[ \Big( 4 a_{3}-\frac{c_{2}}{2}\Big)N^{\alpha(1)}g_{\mu\nu}+\Big(  a_{5}+\frac{c_{2}}{2}\Big)N^{\alpha(2)}g_{\mu\nu}+a_{5}N^{\alpha(3)}g_{\mu\nu}\right]  \nonumber \\
-\frac{1}{\sqrt{-g}}	\hat{\nabla}_{(\mu}  \left[ \Big( 2 a_{5}-\frac{c_{3}}{2}\Big)N_{\nu)}^{(1)}+\Big(  2 a_{4}+\frac{c_{3}}{2}\Big)N_{\nu)}^{(2)}+2 a_{4}N_{\nu)}^{(3)}\right]  \nonumber \\
-\Big( 2 a_{1}-\frac{c_{1}}{2}\Big) N^{\alpha\beta}_{\;\;\;\;\mu}N_{\alpha\beta\nu}-\Big( 2 a_{1}-\frac{c_{1}}{2}\Big) N^{\alpha\beta}_{\;\;\;\;\mu}N_{\beta\alpha\nu}-\frac{c_{1}}{2}N^{\alpha\beta}_{\;\;\;\;(\mu}(N_{\beta|\nu)\alpha}+N_{\alpha|\nu)\beta} ) \nonumber \\
+(N^{\alpha \;\;\beta}_{\;\;(\nu}+N_{(\nu}^{\;\; \alpha\beta})\left[ 2 a_{1}N_{\alpha|\mu)\beta}+ a_{2}N_{\beta|\mu)\alpha}+\Big( 2 a_{1}-\frac{c_{1}}{2}\Big)N_{\mu)\alpha\beta}+\Big( a_{2}+\frac{c_{1}}{2}\Big) N_{\mu)\beta\alpha} \right] \nonumber \\
-\frac{1}{4}(N_{\beta(\nu|\alpha}-N_{\beta\alpha(\nu})\Big( 2 b_{1}(N^{\beta \;\;\alpha}_{\;\;\mu)}-b_{2}N^{\alpha \;\;\beta}_{\;\;\mu)}-2 b_{1}N^{\beta\alpha}_{\;\;\;\mu)}+ b_{2}N^{\alpha\beta}_{\;\;\;\mu)}\Big)+b_{1}N_{\mu\alpha\beta}N_{\nu}^{\;\;[\alpha\beta]} \nonumber \\
+(N_{\mu\nu\alpha}+N_{\nu\mu\alpha})\left[ \Big( 4 a_{3}-\frac{c_{2}}{2}\Big) N^{(1)\alpha}+\Big( a_{5}+\frac{c_{2}}{2} \Big) N^{(2)\alpha}+ a_{5}N^{(3)\alpha} \right] \nonumber \\
-\Big( 4 a_{3}-\frac{c_{3}}{2}\Big) N_{\mu}^{(1)}N_{\nu}^{(1)}-\Big( \frac{b_{3}}{4}-a_{4}\Big)  N_{\mu}^{(2)}N_{\nu}^{(2)}+a_{4} N_{\mu}^{(3)}N_{\nu}^{(3)}
-\Big( c_{3}-\frac{b_{3}}{2}\Big)  N_{(\mu}^{(1)}N_{\nu)}^{(2)}+ 2 a_{4}  N_{(\mu}^{(2)}N_{\nu)}^{(3)} \label{GReff}
\end{gather}
and recall that 
\beq
N_{\alpha\mu\nu}=\kappa\Big( \tilde{a}_{11}	\hat{\Delta}_{\alpha\mu\nu}+\tilde{a}_{12}\hat{\Delta}_{\nu\alpha\mu}+\tilde{a}_{13}\hat{\Delta}_{\mu\nu\alpha}+\tilde{a}_{14}\hat{\Delta}_{\alpha\nu\mu}
+\tilde{a}_{15} \hat{\Delta}_{\nu\mu\alpha}+\tilde{a}_{16} \hat{\Delta}_{\mu\alpha\nu}\Big) 
\eeq
That is, we see that the Theory is GR with modified/extended matter sources. Indeed, since the distortion is linear in the hypermomentum the extra terms appearing on the right-hand side of ($\ref{GReff}$) are quadratic on the hypermomentum and linear in its derivatives. As a result, the source now is  not solely given by the  usual energy-momentum tensor of matter but we also have hypermomentum contributions. In this sense the microstructure of matter (hypermomentum) modifies the usual matter content in a highly non-trivial manner. Of course, given the complexity of ($\ref{GReff}$) when all $11$ parameters are included, no concrete conclusions can be drawn and the exact effects of these contributions depend both on the number of parameters included and on the given model for the hypermomentum source that one considers.

\section{Generalized Quadratic Theory}

It is worth stressing out that in applying the results of $Theorem-1$ in order to solve equation ($\ref{Neq}$) for the affine connection, nowhere did we assume that the coefficients $a_{i}$, $b_{j}$ and $c_{k}$ are constants. Subsequently, the main result for the affine connection will continue to hold true even if we  promote the aforementioned constants to be spacetime functions. In particular, we have the following result.

\begin{theo}
Let $\Phi$ be a collection of matter fields (collectively denoted) and $d\Phi$ their exterior derivatives. Furthermore, let $a_{i}(\Phi, d \Phi)=a_{i}(\Phi, \partial \Phi)$, $b_{j}(\Phi, \partial \Phi)$ and $c_{j}(\Phi, \partial \Phi)$ $i=1,2,...,5$, $j=1,2,3$ be arbitrary functions continuous and differentiable in their arguments\footnote{These functions can just as well depend also on the metric tensor and its derivatives, however not on the connection.}. Then, for the general class of Theories given by the action
\begin{gather}
S[g, \Gamma, \Phi]
=\frac{1}{2 \kappa}\int d^{n}x \sqrt{-g} \Big[  R+ 
b_{1}(\Phi, \partial \Phi)S_{\alpha\mu\nu}S^{\alpha\mu\nu} +
b_{2}(\Phi, \partial \Phi)S_{\alpha\mu\nu}S^{\mu\nu\alpha} +
b_{3}(\Phi, \partial \Phi)S_{\mu}S^{\mu} \nonumber \\
a_{1}(\Phi, \partial \Phi)Q_{\alpha\mu\nu}Q^{\alpha\mu\nu} +
a_{2}(\Phi, \partial \Phi)Q_{\alpha\mu\nu}Q^{\mu\nu\alpha} +
a_{3}(\Phi, \partial \Phi)Q_{\mu}Q^{\mu}+
a_{4}(\Phi, \partial \Phi)q_{\mu}q^{\mu}+
a_{5}(\Phi, \partial \Phi)Q_{\mu}q^{\mu} \nonumber \\
+c_{1}(\Phi, \partial \Phi)Q_{\alpha\mu\nu}S^{\alpha\mu\nu}+
c_{2}(\Phi, \partial \Phi)Q_{\mu}S^{\mu} +
c_{3}(\Phi, \partial \Phi)q_{\mu}S^{\mu} 
\Big] +S_{M}[g, \Gamma, \Phi] \label{genquad}
\end{gather}	
	the affine connection is given by
	\beq
	\Gamma^{\lambda}_{\;\;\;\mu\nu}=\tilde{\Gamma}^{\lambda}_{\;\;\;\mu\nu}+\kappa\Big( \tilde{a}_{11}	\hat{\Delta}_{\alpha\mu\nu}+\tilde{a}_{12}\hat{\Delta}_{\nu\alpha\mu}+\tilde{a}_{13}\hat{\Delta}_{\mu\nu\alpha}+\tilde{a}_{14}\hat{\Delta}_{\alpha\nu\mu}
	+\tilde{a}_{15} \hat{\Delta}_{\nu\mu\alpha}+\tilde{a}_{16} \hat{\Delta}_{\mu\alpha\nu}\Big) 
	\eeq
	where now $\tilde{\alpha}_{ij}=\tilde{\alpha}_{ij}(\Phi, \partial \Phi)$ and $\hat{\Delta}_{\mu\nu\alpha}$ is given by ($\ref{Dhat}$).

	\end{theo}

\textit{Proof.}	Following an identical procedure with that of section $VI$ and by simply promoting the $a_{i}'s, b_{j}'s$ and $c_{j}'s$ to be functions of $\Phi$ and $d\Phi$ we arrive at the stated result. 

\textbf{Comment.} Note that in the action ($\ref{genquad}$) are included Scalar-Tensor and Vector-Tensor Theories as special cases. We have therefore provided the exact form of the affine connection for Scalar-Tensor and Vector-Tensor Theories whose actions are of the same  form as $(\ref{genquad})$.

\section{Conclusions}

We have considered the most general parity even quadratic MAG Theory in $n$ dimensions. The aforementioned Theory consists of the usual Einstein-Hilbert term along with the $11$ parity even quadratic terms in torsion and non-metricity. Additionally, in order to keep things fairly general, we imposed no restriction on the matter sector, allowing it to have an arbitrary hypermomentum. Then, after deriving the field equations by varying with respect to the metric and the independent affine connection,  we first focused on the latter connection field equations. By using a recent result on how to solve such linear but complicated tensor equations (see cf. \cite{Iosifidis:2021ili}) we were able to find the exact solution for the affine connection in terms of the (arbitrary) hypermomentum and its traces. Given two fairly general non-degeneracy conditions, the uniqueness of our exact solution is also guaranteed. With this result we also extracted the exact forms of torsion and non-metricity for the general Theory.

 Having obtained the exact form of the connection, we then used a post-Riemannian expansion on the metric-field equations and expressed the latter in an effective GR-like form with modified/extended source terms that depend quadratically on the hypermomentum and linearly on its derivative.  These extra terms come on top of the usual energy-momentum tensor of matter and modify the matter content in a highly non-trivial manner, as seen from equation ($\ref{GReff}$). Furthermore, the non-vanishing of the hypermomentum has a close relation with the intrinsic properties of matter \cite{Puetzfeld:2007ye}. With this in mind, we see that the additional modifications compared to GR have a very nice and clear physical interpretation, namely, the microstructure of matter modifies the gravitational field. However, the exact  effect of this microstructure in the general case where all $11$ parameters are included is not so easy to decode given the complexity of ($\ref{GReff}$). Nevertheless, more manageable expressions and subsequently a better understanding of these additional hypermomentum terms can be acquired once a specific choice for the matter sector is made.  It is also important to note that, just like in the Einstein-Cartan Theory, once the sources are switched off (i.e. vacuum case) the complicated quadratic Theory reduces to GR. In other words, the new interactions are tied to matter (i.e. they do not propagate) and as a result outside  matter the quadratic Theory is indistinguishable from GR but changes considerably when matter with intrinsic structure is considered.

Going one step further we considered the general class of quadratic Metric-Affine Theories as given by ($\ref{genquad}$) where now the coupling constants between the torsion and non-metricity invariants have been promoted to functions (of some given matter fields). Employing the same procedure with the case of constant coefficients we were able to solve again for the affine connection. In this generalization specific Scalar-Tensor and Vector-Tensor Metric-Affine Theories are also included. Therefore, our results here will be of prominent importance in analysing such Theories. Another very interesting application would be to compute the right hand-side of ($\ref{GReff}$) for specific matter types (i.e given $T_{\mu\nu}$ and $\Delta_{\alpha\mu\nu}$ tensors) and see what the new interactions look like especially in comparison with the Einstein-Cartan Theory.

	\section{Acknowledgments}	This research is co-financed by Greece and the European Union (European Social Fund- ESF) through the
Operational Programme 'Human Resources Development, Education and Lifelong Learning' in the context
of the project “Reinforcement of Postdoctoral Researchers - 2
nd Cycle” (MIS-5033021), implemented by the
State Scholarships Foundation (IKY).

\appendix
\section{Relations among parameters}
\begin{gather}
\alpha_{1}=4 a_{1}+b_{1}-c_{1}\;\;, \;\; \alpha_{2}=\alpha_{3}=-1+ 2 a_{2}+\frac{b_{2}+c_{1}}{2} \nonumber \\
\alpha_{4}=2 a_{2}-b_{1}+c_{1}\;\;, \;\; \alpha_{5}=2 a_{2}-\frac{b_{2}}{2} \;\;, \;\; \alpha_{6}= 4 a_{1}-\frac{b_{2}}{2}-c_{1} \nonumber \\
\alpha_{71}=2 a_{5}+c_{2}-\frac{b_{3}+c_{3}}{2} \;\;, \;\; \alpha_{72}=2 a_{4}+c_{3}+\frac{b_{3}}{2} \;\;, \;\; \alpha_{73}= 1+ 2a_{4}+\frac{c_{3}}{2} \nonumber \\
\alpha_{81}=8 a_{3}- 2 c_{2}+\frac{b_{3}}{2} \;\;, \;\; \alpha_{82}= 2 a_{5}+c_{2}-\frac{c_{3}+b_{3}}{2}\;\;, \;\; \alpha_{83}= 2 a_{5}-\frac{c_{3}}{2} \nonumber \\
\alpha_{91}=2 a_{5}-\frac{c_{3}}{2} \;\;, \;\; \alpha_{92}= 2 a_{4}+\frac{c_{3}}{2}+1 \;\;, \;\; \alpha_{93}=2 a_{4}
\end{gather}

\subsection{Elements of matrix $A$ in terms of Theory parameters}

\subsection{The $\gamma's$}
The relations between the elements of $\Gamma$ and the parameters $a_{i}$ read
\begin{gather}
\gamma_{11}=a_{1}+a_{3}+a_{71}+n a_{81}+a_{91}\;\;, \;\; \gamma_{12}=a_{2}+a_{4}+a_{72}+n a_{82}+ a_{92}\;\;, \;\; \gamma_{13}=a_{5}+a_{6}+a_{73}+n a_{83}+a_{93} \nonumber \\
\gamma_{21}=a_{2}+a_{5}+n a_{71}+ a_{81}+a_{91}\;\;, \;\; \gamma_{22}=a_{1}+a_{6}+n a_{72}+a_{82}+ a_{92}\;\;, \;\; \gamma_{23}=a_{3}+a_{4}+n a_{73}+ a_{83}+a_{93} \nonumber \\		
\gamma_{31}=a_{5}+a_{6}+a_{71}+ a_{81}+ n a_{91}\;\;, \;\; \gamma_{32}=a_{3}+a_{4}+a_{72}+ a_{82}+n a_{92}\;\;, \;\; \gamma_{31}=a_{1}+a_{2}+a_{73}+ a_{83}+n a_{93} \nonumber
\end{gather}

\section{Post-Riemannian Expansions}

\subsection{Distortion vectors in terms of torsion and non-metricity vectors}
Contracting ($\ref{QNSN}$) in all possible ways with the metric  we find
\beq
N_{(1)}^{\mu}=\frac{1}{2}Q^{\mu}\; \;, \;\; N_{(2)}^{\mu}=\frac{1}{2}Q^{\mu}+2 S^{\mu}\;\;, \;\; N_{(3)}^{\mu}=q^{\mu}-\frac{1}{2}Q^{\mu}-2 S^{\mu}
\eeq
which relate the distortion vectors in terms of those of torsion and non-metricity. Inverting them we have
\beq
Q^{\mu}=2 N_{(1)}^{\mu}\;\;, \;\; q^{\mu}=N_{(2)}^{\mu}+N_{(3)}^{\mu}\;\;, \;\; S^{\mu}=\frac{1}{2}\Big( N_{(2)}^{\mu}-N_{(1)}^{\mu}\Big)
\eeq
Then, using again ($\ref{QNSN}$) along with the above relations and the definition of the Palatini tensor we can express the latter solely in terms of the distortion as
\beq
P^{\alpha\mu\nu}=g^{\mu\nu}N_{(2)}^{\alpha}+g^{\nu\alpha}N_{(3)}^{\mu}-(N^{\mu\nu\alpha}+N^{\nu\alpha\mu})
\eeq

\subsection{Post-Riemannian expansions}
The Ricci tensor is expanded as
\begin{equation}
R_{\nu\beta}=\tilde{R}_{\nu\beta}+ \tilde{\nabla}_{\mu}N^{\mu}_{\;\;\;\nu\beta}-\tilde{\nabla}_{\beta}N^{(2)}_{\nu}+N^{(2)}_{\lambda}N^{\lambda}_{\;\;\;\nu\beta}-N^{\mu}_{\;\;\;\rho\beta}N^{\rho}_{\;\;\;\nu\mu}
\end{equation}
and the Ricci scalar as
\begin{equation}
R=\tilde{R}+ \tilde{\nabla}_{\mu}( N^{(3)\mu}-N^{(2)\mu})+ N^{(3)}_{\mu}N^{(2)\mu}-N_{\alpha\mu\nu}N^{\mu\nu\alpha} \label{Recomp}
\end{equation}
As a result for the Einstein-like combination we find
\begin{gather}
R_{(\mu\nu)}-\frac{R}{2}g_{\mu\nu}=	\Big( \tilde{R}_{(\mu\nu)}-\frac{\tilde{R}}{2}g_{\mu\nu}\Big)+\tilde{\nabla}_{\alpha}N^{\alpha}_{\;\;\;(\mu\nu)}-\tilde{\nabla}_{(\nu}N^{(2)}_{\mu)}+N^{(2)}_{\lambda}N^{\lambda}_{\;\;\;(\mu\nu)}-N^{\alpha}_{\;\;\;\lambda(\nu}N^{\lambda}_{\;\;\;\mu)\alpha}\nonumber \\
-\frac{1}{2}g_{\mu\nu}\Big[ \tilde{\nabla}_{\alpha}( N^{(3)\alpha}-N^{(2)\alpha})+N^{(3)}_{\alpha}N^{(2)\alpha}-N_{\alpha\beta\gamma}N^{\beta\gamma\alpha} \Big]
\end{gather}
Continuing we calculate
\begin{gather}
b_{1}S_{\alpha\mu\nu}S^{\alpha\mu\nu} +
b_{2}S_{\alpha\mu\nu}S^{\mu\nu\alpha} +
a_{1}Q_{\alpha\mu\nu}Q^{\alpha\mu\nu} +
a_{2}Q_{\alpha\mu\nu}Q^{\mu\nu\alpha} +
+c_{1}Q_{\alpha\mu\nu}S^{\alpha\mu\nu}= \nonumber \\
\Big( \frac{b_{1}}{2}+ 2 a_{1}-c_{1}\Big) N_{\alpha\mu\nu}N^{\alpha\mu\nu}+	\Big( -\frac{b_{1}}{2}+  a_{2}+c_{1}\Big)N_{\alpha\mu\nu}N^{\alpha\nu\mu} \nonumber \\
+	\Big( \frac{b_{2}}{2}+ 2 a_{2}+c_{1}\Big)N_{\alpha\mu\nu}N^{\mu\nu\alpha}+	\Big( -\frac{b_{2}}{2}+ 2 a_{1}+a_{2}-c_{1}\Big)N_{\mu\nu\alpha}N^{\nu\mu\alpha}
\end{gather}
and also
\begin{gather}
b_{3}S_{\mu}S^{\mu} +
a_{3}Q_{\mu}Q^{\mu}+
a_{4}q_{\mu}q^{\mu}+
a_{5}Q_{\mu}q^{\mu}+
c_{2}Q_{\mu}S^{\mu} +
c_{3}q_{\mu}S^{\mu}= \nonumber \\
=\Big( \frac{b_{3}}{4}+4 a_{3}-c_{2} \Big)N^{(1)}_{\mu}N^{(1)}_{\nu}g^{\mu\nu}+\Big( \frac{b_{3}}{4}+a_{4}+\frac{c_{3}}{4} \Big)N^{(2)}_{\mu}N^{(2)}_{\nu}g^{\mu\nu}+a _{4}N^{(3)}_{\mu}N^{(3)}_{\nu}g^{\mu\nu} \nonumber \\
+\Big( 2 a_{5}-\frac{c_{3}}{2}+c_{2}-\frac{b_{3}}{2}\Big) N^{(1)}_{\mu}N^{(2)}_{\nu}g^{\mu\nu}	+\Big( 2 a_{4}+\frac{c_{3}}{2}\Big)N^{(2)}_{\mu}N^{(3)}_{\nu}g^{\mu\nu}+\Big( 2 a_{5}-\frac{c_{3}}{2}\Big) N^{(1)}_{\mu}N^{(3)}_{\nu}g^{\mu\nu}
\end{gather}
such that
\begin{gather}
\mathcal{L}_{2}=\Big( \frac{b_{1}}{2}+ 2 a_{1}-c_{1}\Big) N_{\alpha\mu\nu}N^{\alpha\mu\nu}+	\Big( -\frac{b_{1}}{2}+  a_{2}+c_{1}\Big)N_{\alpha\mu\nu}N^{\alpha\nu\mu} \nonumber \\
+	\Big( \frac{b_{2}}{2}+ 2 a_{2}+c_{1}\Big)N_{\alpha\mu\nu}N^{\mu\nu\alpha}+	\Big( -\frac{b_{2}}{2}+ 2 a_{1}+a_{2}-c_{1}\Big)N_{\mu\nu\alpha}N^{\nu\mu\alpha} \nonumber \\
=\Big( \frac{b_{3}}{4}+4 a_{3}-c_{2} \Big)N^{(1)}_{\mu}N^{(1)}_{\nu}g^{\mu\nu}+\Big( \frac{b_{3}}{4}+a_{4}+\frac{c_{3}}{4} \Big)N^{(2)}_{\mu}N^{(2)}_{\nu}g^{\mu\nu}+a _{4}N^{(3)}_{\mu}N^{(3)}_{\nu}g^{\mu\nu} \nonumber \\
+\Big( 2 a_{5}-\frac{c_{3}}{2}+c_{2}-\frac{b_{3}}{2}\Big) N^{(1)}_{\mu}N^{(2)}_{\nu}g^{\mu\nu}	+\Big( 2 a_{4}+\frac{c_{3}}{2}\Big)N^{(2)}_{\mu}N^{(3)}_{\nu}g^{\mu\nu}+\Big( 2 a_{5}-\frac{c_{3}}{2}\Big) N^{(1)}_{\mu}N^{(3)}_{\nu}g^{\mu\nu}
\end{gather}
Finally, we calculate
\begin{gather}
W_{\alpha(\mu\nu)}+\Pi_{\alpha(\mu\nu)}=\Big( 4 a_{1}-\frac{c_{1}}{2}\Big) N_{(\mu\nu)\alpha}+\Big( 2 a_{2}+\frac{c_{1}}{2}\Big)N_{(\mu|\alpha|\nu)}+2 a_{2}N_{\alpha(\mu\nu)} \nonumber \\
\left[ \Big( 4 a_{3}-\frac{c_{2}}{2}\Big)N_{\alpha}^{(1)}+\Big(  a_{5}+\frac{c_{2}}{2}\Big)N_{\alpha}^{(2)}+a_{5}N_{\alpha}^{(3)}\right] g_{\mu\nu} \nonumber \\
+	g_{\alpha(\mu}  \left[ \Big( 2 a_{5}-\frac{c_{3}}{2}\Big)N_{\nu)}^{(1)}+\Big(  2 a_{4}+\frac{c_{3}}{2}\Big)N_{\nu)}^{(2)}+2 a_{4}N_{\nu)}^{(3)}\right]  
\end{gather}

as well as 
\begin{gather}
A_{(\mu\nu)}+B_{(\mu\nu)}+C_{(\mu\nu)}=Q_{(\mu}^{\;\;\;\;\alpha\beta}\Big( a_{1}Q_{\nu)\alpha\beta}+c_{1} S_{\nu)\alpha\beta}\Big) -\Big( 2 a_{1}Q_{\alpha\beta(\mu}+c_{1} S_{\alpha\beta(\mu}+ a_{2} Q_{\beta\alpha(\mu}\Big) Q^{\alpha\beta}_{\;\;\;\;\nu)}\nonumber \\
+\Big( 2 b_{1}S_{(\mu}^{\;\;\;\alpha\beta}-b_{2} S_{(\mu}^{\;\;\;\beta\alpha}\Big) S_{\nu)\alpha\beta}-b_{1} S_{\alpha\beta\mu}S^{\alpha\beta}_{\;\;\;\nu}-Q_{\alpha\mu\nu}\Big( 2 a_{3}Q^{\alpha}+a_{5}q^{\alpha}+c_{2} S^{\alpha} \Big) \nonumber \\
+a_{3}Q_{\mu}Q_{\nu}-a _{4}q_{\mu}q_{\nu}+b_{3}S_{\mu}S_{\nu}+c_{2}S_{(\mu}Q_{\nu)}= \nonumber \\
=\Big( 2 a_{1}-\frac{c_{1}}{2}\Big) N^{\alpha\beta}_{\;\;\;\;\mu}N_{\alpha\beta\nu}+\Big( 2 a_{1}-\frac{c_{1}}{2}\Big) N^{\alpha\beta}_{\;\;\;\;\mu}N_{\beta\alpha\nu}+\frac{c_{1}}{2}N^{\alpha\beta}_{\;\;\;\;(\mu}(N_{\beta|\nu)\alpha}+N_{\alpha|\nu)\beta} ) \nonumber \\
-(N^{\alpha \;\;\beta}_{\;\;(\nu}+N_{(\nu}^{\;\; \alpha\beta})\left[ 2 a_{1}N_{\alpha|\mu)\beta}+ a_{2}N_{\beta|\mu)\alpha}+\Big( 2 a_{1}-\frac{c_{1}}{2}\Big)N_{\mu)\alpha\beta}+\Big( a_{2}+\frac{c_{1}}{2}\Big) N_{\mu)\beta\alpha} \right] \nonumber \\
+\frac{1}{4}(N_{\beta(\nu|\alpha}-N_{\beta\alpha(\nu})\Big( 2 b_{1}(N^{\beta \;\;\alpha}_{\;\;\mu)}-b_{2}N^{\alpha \;\;\beta}_{\;\;\mu)}-2 b_{1}N^{\beta\alpha}_{\;\;\;\mu)}+ b_{2}N^{\alpha\beta}_{\;\;\;\mu)}\Big)-b_{1}N_{\mu\alpha\beta}N_{\nu}^{\;\;[\alpha\beta]} \nonumber \\
-(N_{\mu\nu\alpha}+N_{\nu\mu\alpha})\left[ \Big( 4 a_{3}-\frac{c_{2}}{2}\Big) N^{(1)\alpha}+\Big( a_{5}+\frac{c_{2}}{2} \Big) N^{(2)\alpha}+ a_{5}N^{(3)\alpha} \right] \nonumber \\
+\Big( 4 a_{3}-\frac{c_{3}}{2}\Big) N_{\mu}^{(1)}N_{\nu}^{(1)}+\Big( \frac{b_{3}}{4}-a_{4}\Big)  N_{\mu}^{(2)}N_{\nu}^{(2)}-a_{4} N_{\mu}^{(3)}N_{\nu}^{(3)}\nonumber \\
+\Big( c_{3}-\frac{b_{3}}{2}\Big)  N_{(\mu}^{(1)}N_{\nu)}^{(2)}- 2 a_{4}  N_{(\mu}^{(2)}N_{\nu)}^{(3)}
\end{gather}

	\bibliographystyle{unsrt}
\bibliography{ref}

\end{document}